\documentclass[reprint,nofootinbib,twocolumn
]{revtex4-2}

\usepackage{graphicx}
\usepackage{dcolumn}
\usepackage{bm}

\usepackage{amsmath}
\usepackage{amssymb}
\usepackage{graphicx}
\usepackage{bbm}
\usepackage{color}
\usepackage{url}
\usepackage{multirow}
\usepackage{hyperref}
\usepackage[table,xcdraw]{xcolor}
\usepackage{enumerate}
\usepackage[normalem]{ulem}
\usepackage{comment}

\definecolor{deepfuchsia}{rgb}{0.76, 0.33, 0.76}

\newcommand{\SO}[1]{\ensuremath{\mathrm{SO}(#1)}}

\newcommand{\vev}[1]{\langle #1 \rangle}
\newcommand{\be}{\begin{equation}}  
\newcommand{\ee}{\end{equation}}
\newcommand{\into}{\ensuremath{\,\rightarrow\,}}
\newcommand{\MP}{\ensuremath{M_\mathrm{Pl}}}

\begin{document}

\preprint{APS/123-QED}

\title{\textbf{No room for minimal monopole dark matter} 
}

\author{Felix Br\"ummer$^1$}

\author{Giacomo Ferrante$^{1,2}$}
\email{Contact author: giacomo.ferrante@ulb.be}

\author{Theodore Fischer$^{1,3}$}

\author{Michele Frigerio$^{3,4}$}

\affiliation{$^1$ Laboratoire Univers et Particules de Montpellier (LUPM), University of Montpellier \& CNRS, Montpellier, France}
\affiliation{$^2$ Service de Physique Théorique, Université Libre de Bruxelles (ULB), Brussels, Belgium}
\affiliation{$^3$ Laboratoire Charles Coulomb (L2C), University of Montpellier \& CNRS, Montpellier, France\\}
\affiliation{$^4$ Laboratoire de Physique Théorique et Hautes Énergies (LPTHE),
Sorbonne Université \& CNRS, Paris, France\\}

\date{\today}

\begin{abstract}
The magnetic monopole of a dark sector has been advocated as an appealing dark matter candidate. We revisit the computation of the monopole abundance $\Omega_M$, generated by a thermal phase transition in the minimal 't Hooft-Polyakov model.
We explore the three regimes where the phase transition is second order, weakly first order, or supercooled, identifying the parameter space regions where $\Omega_M$ can match the observed dark matter abundance.
However, the dark sector necessarily contains a stable electrically-charged particle, namely a massive vector boson, with a calculable abundance $\Omega_{W'}$.
We show that, 
under minimal assumptions,
$\Omega_{W'}$ is always far larger than $\Omega_M$:
dark monopoles cannot constitute a sizeable fraction of dark matter. 
\end{abstract}

\maketitle

\section{Introduction}

A gauge interaction is said to belong to a dark sector if none of the Standard Model (SM) particles are charged under it. Dark-sector gauge symmetries may be broken by the Higgs mechanism, and may thus give rise to magnetic monopoles. These are extended topological defects, consisting of nontrivial gauge-field and Higgs-field configurations. At low energies, they effectively behave as massive classical particles.

Since monopoles are topologically stable, they will contribute to the dark matter (DM) of the universe. Historically, this was regarded as a problem for grand-unified theories, as the breaking of the unified gauge group to the SM gauge group would abundantly produce superheavy magnetic monopoles of ordinary electromagnetism, and thus overclose the universe. 
If, however, the monopoles are part of a dark sector, the symmetry breaking scale can be much lower than the unification scale. Then the monopole abundance need not lead to overclosure. In fact, being stable massive particles, dark-sector monopoles might well account for all or part of the observed DM.

The production mechanisms for monopole DM are quite different from ordinary particle DM. This makes it an interesting object for study. The possibility that all of the DM consists of monopoles, produced via a thermal phase transition in the early universe, has been analysed in \cite{Murayama:2009nj,GomezSanchez:2011orv,Baek:2013dwa,Khoze:2014woa,Kawasaki:2015lpf,Sato:2018nqy,Daido:2019tbm,Graesser:2020hiv,Yang:2022quy}. 
Monopoles produced during the stage of preheating have been studied in \cite{Bai:2020ttp}.

The minimal dark-sector model featuring monopoles with calculable properties has an $\SO{3}$ gauge group, broken to $\SO{2}$ by a Higgs triplet vacuum expectation value \cite{Georgi:1972cj}, leading to 't Hooft-Polyakov monopoles \cite{tHooft:1974kcl, Polyakov:1974ek}. In this model there exists another stable DM candidate, namely a massive $W'$ gauge boson, which is the lightest state carrying electric charge under the unbroken symmetry. Indeed, in any model featuring a magnetic monopole, the lightest electrically-charged state will be stable as well. It is an interesting question whether the DM abundance can be dominated by dark-sector monopoles, rather than by elementary particles with dark electric charge.

In this Letter, we will show that the answer to this question is negative for all of the parameter space of the minimal 't Hooft-Polyakov monopoles, produced during a thermal phase transition. With rather generic assumptions (essentially, we demand that the dark-sector couplings are in the perturbative regime, and the interactions with the SM play a subleading role), we find that the thermal $W'$ abundance always exceeds the monopole abundance by far.

\section{Minimal dark-sector monopoles}\label{ssec:Model}

Let $G=\SO{3}$ be a dark-sector gauge group and $\phi$ be a real scalar $\SO{3}$ triplet. The most general renormalizable potential for $\phi$ reads
\begin{equation}\label{eq:Pot}
    V(\phi) = -\frac{\mu^2}{2} \phi^2 + \frac{\lambda}{4} (\phi^2)^2+\frac{\lambda_{\phi H}}{2}\,\phi^2\,|H|^2\,,
\end{equation}
with $H$ the SM Higgs doublet. For $\lambda>0$ and $\mu^2>0$, and neglecting the Higgs portal term for the time being, the potential is minimized at 
$\vev{\phi^2}=\mu^2/\lambda \equiv\eta^2$,
and $\SO{3}$ is broken to $\SO{2}$. There is a massive scalar radial mode $\rho$ with $m_\rho^2 = 2\lambda \eta^2$, and two 
massive gauge bosons $W'^\pm$ with $m_{W'}^2 = g^2 \eta^2$, where $g$ is the dark gauge coupling.  A third vector boson $\gamma'$, associated to the unbroken $\SO{2}$, remains massless.

The second homotopy group of the vacuum manifold is nontrivial, $\pi_2\left[\SO{3}/\SO{2}\right] = \mathbb{Z}$. The model therefore features stable monopole configurations, famously constructed in \cite{tHooft:1974kcl, Polyakov:1974ek}.
The monopole $M$ with unit winding number has mass $m_M \approx  4\pi \eta/g$, magnetic charge $q_M = 4\pi/g$, no electric charge (for minimality, we assume a vanishing theta term for the dark gauge fields), and spin zero.  The monopole core radius $r_M$ scales as $r_M \approx (g \eta)^{-1}$.
For perturbatively small values of $g$,  $r_M$ is much larger than the monopole Compton wavelength $\lambda_M\equiv 1/m_M \le (g^2/4\pi) (g\eta)^{-1}$; therefore the monopole can be treated as a classical object.

\section{Phase transitions}

We will assume that, in the early universe, the dark sector and the SM are thermalised via the Higgs portal coupling $\lambda_{\phi H}$, and that the reheating temperature is larger than the $\SO{3}$-symmetry breaking scale $\eta$. Thermal effects then correct the zero-temperature potential in Eq.~\eqref{eq:Pot} and lead to symmetry restoration. As the universe cools down, a symmetry-breaking phase transition occurs at some critical temperature $T_c$. The universe eventually settles in the symmetry-breaking vacuum, either smoothly in the case of a second-order phase transition (SOPT), or via bubble nucleation in the case of a first-order phase transition (FOPT). In either case, dark monopoles are produced.

\begin{figure}[tb]
\begin{center}
\includegraphics[width=.5\textwidth]{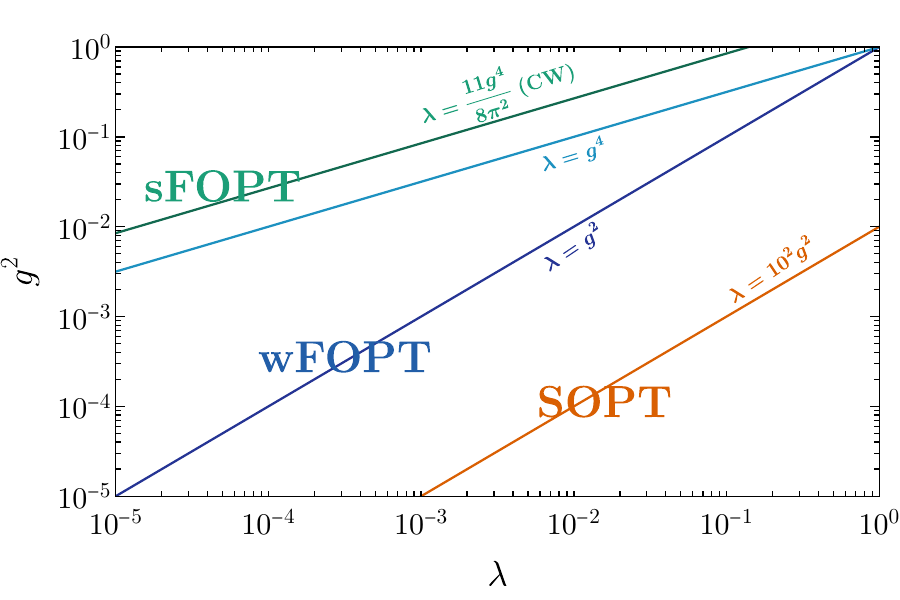}
    \caption{
   The nature of the phase transition depends on the choice of the couplings $\lambda$ and $g^2$. Above the dark-blue line, 
   finite-temperature perturbation theory holds.
   Below the light-blue line,  high-temperature expansion holds.  
   The green line indicates the Coleman-Weinberg scenario. 
   The orange line indicates a benchmark for a SOPT.
   }
\label{fig:PhDiag}  
\end{center}
\end{figure}

We assume that the couplings $g$ and $\lambda$ are perturbatively small, and that $\lambda_{\phi H}$ is small enough not to significantly affect the dark-sector thermal effective potential.
The nature of the phase transition depends on the ratio $\lambda/g^2$, as sketched in Fig.~\ref{fig:PhDiag}.

\begin{itemize}
\item For $\lambda\ll g^2$, finite-temperature perturbation theory can be used to establish that the phase transition is of the first order \cite{Dolan:1973qd}. If, moreover, $\lambda\gg g^4$, then the one-loop thermal effective potential close to $T_c$,
\be\label{eq:Veff}
\qquad V_{\rm eff}(\phi,T)\approx\frac{m^2(T)}{2}\phi^2-\frac{\delta(T)}{3}|\phi|^3+\frac{\lambda(T)}{4}(\phi^2)^2,
\ee
can be computed in a high-temperature expansion  \cite{Dolan:1973qd, Anderson:1991zb, Dine:1992wr, Arnold:1992rz}. The phase transition in this region is found to be \emph{weakly} first-order, since bubbles start nucleating at temperatures immediately below $T_c$.
\item For $\lambda < g^4$, perturbation theory can still be used but the high-temperature expansion of the effective potential is unreliable. This includes e.g.~the Coleman-Weinberg scenario, where the second derivative of the $T=0$ one-loop effective potential at $\phi=0$ is renormalized to zero. This implies that symmetry breaking is radiatively induced, and that $\lambda=11 g^4/8\pi^2$, where $\lambda$ is the renormalized coupling in the symmetry-breaking vacuum. The phase transition is \emph{strongly} first-order, as the tunneling rate to the true vacuum at $T\lesssim T_c$ is still highly suppressed. Bubbles of true vacuum will only form at a much lower nucleation temperature, $T_n\ll T_c$. Such phase transitions are dubbed \emph{supercooled}.
\item For $g^2<\lambda$, perturbation theory cannot be used to establish the nature of the phase transition. Lattice studies of the closely related electroweak (EW) phase transition in the SM \cite{Karsch:1996yh,Gurtler:1997hr,Rummukainen:1998as} and of the Abelian Higgs model \cite{Kajantie:1997hn, Hove:2000rzx} indicate that it turns into a crossover at some critical ratio $\left(g^2/\lambda\right)_{\rm crit}\sim{\cal O}(0.1)$ in the nonperturbative region. It is reasonable to expect that the model we are studying will behave similarly. In the limit $g\into 0$ of a global $\SO{3}$ symmetry, the phase transition is of the second order, since by dimensional reduction \cite{Ginsparg:1980ef} the theory can be mapped to the ${\rm O}(3)$ model in three Euclidean dimensions \cite{Zinn-Justin:2002ecy}. By continuity, the phase transition should also be of the second order ``for all practical purposes'' for nonzero but sufficiently small $g$. 
 \end{itemize}

As the phase transition takes place, the order parameter  changes from $\langle \phi \rangle = 0$ to a non-zero value. While the absolute value of $\langle \phi \rangle$ is unequivocally determined by energy minimisation, its orientation in field space is not: the field takes random orientations in the vacuum manifold on scales larger than the field correlation length $\xi$. Therefore, the universe is fragmented into domains of typical length $\xi$, each characterised by a different orientation of $\langle\phi\rangle$. At the intersection of these domains, a monopole may form with a probability $p$ which depends on the topology of the vacuum manifold. Therefore, the monopole number density can be estimated as $n_M \approx p\,\xi^{-3}$.
In our case, the vacuum manifold is $S^2$ and $p=1/8$ \cite{Kibble:1976sj}.
As we review below, the computation of the correlation length $\xi$ strongly depends on the details of the phase transition.

\section{Monopoles from second-order phase transitions}

During a SOPT, monopoles are formed by the Kibble-Zurek (KZ) mechanism \cite{Kibble:1976sj,Zurek:1985qw}; see e.g.~\cite{Zurek:1996sj, delCampo:2013nla} for reviews. As the temperature approaches the critical temperature $T_c$, both the correlation length and the relaxation time diverge with critical exponents $\nu$ and $\mu$ respectively. Since our model is in the Heisenberg universality class of the O(3) model in three dimensions, we take $\nu\approx 0.7$ \cite{Zinn-Justin:2002ecy}, and since its dispersion relation is relativistic, we have $\mu\approx\nu$ \cite{Zurek:1996sj, Murayama:2009nj} 
Close to $T_c$, the system can no longer equilibrate due to the divergent relaxation time, and fluctuations freeze at a spatial scale
\be
\label{eq:correlation_lgth}
    \xi = H(T_c)^{-1}\, \left[H(T_c)\xi_0\right]^\frac{1}{1+\nu} \equiv \xi_{\rm KZ}\,,
\ee
where  $\xi_0^2\approx 1/m_\rho^2$ and $H$ is the Hubble parameter, $H^2=\gamma_* T^4/\MP^2$, with $\MP\approx 2.4\cdot 10^{18}$ GeV, $\gamma_*\equiv \pi^2 g_*/90$,
and $g_*$ the number of relativistic degrees of freedom.

We define the comoving monopole number density by
$Y_M\equiv n_M/s$, with the entropy density $s=4\gamma_* T^3$.
Using  $n_M \approx \xi^{-3}/8$ and $T_c^2\approx 12\,\eta^2/5$, we obtain from Eq.~\eqref{eq:correlation_lgth}
\be
\begin{aligned}\label{eq:MonoSOPT}
Y_M \approx \frac{1}{32} \left(\frac{5}{6}\lambda\right)^{\frac{3}{2+2\nu}} \gamma_*{}^{\frac{\nu-2}{2+2\nu}}\left(\frac{T_c}{\MP}\right)^{\frac{3\nu}{1+\nu}}\,.
\end{aligned}
\ee
This scaling with $T_c$ matches the generic prediction  \cite{Murayama:2009nj} for monopoles from a SOPT, while the prefactor is specific to the 't Hooft-Polyakov model studied here. 
In Fig.~\ref{fig:SOPT}, the green line labelled `KZ' shows the monopole abundance  obtained from  Eq.~\eqref{eq:MonoSOPT}.

The above discussion is valid as long as the monopoles are effectively point-like. However, in the global limit $g\into 0$, the monopole mass and radius diverge. For $r_M>\xi_{\rm KZ}$, the KZ estimate of the number density must clearly break down. Analytical studies and numerical simulations have shown that global monopoles enter a scaling regime with an ${\cal O}(1)$ number $\zeta$ of monopoles per Hubble volume, $n_M=\zeta H^3$ \cite{Barriola:1989hx, Rhie:1990kc,Yamaguchi:2001xn,Martins:2008zz}\footnote{For possible deviations
from the scaling regime, see the recent simulations of global monopole
networks in \cite{Nakano:2026zme}.}. In radiation domination, \cite{Yamaguchi:2001xn} finds $\zeta= 3.44\pm 0.56$.
This global limit is a good approximation as long as $r_M>1/H$.
To roughly estimate the present-day monopole number density, 
we assume that the scaling regime ends once $r_M\lesssim 1/H$, and that afterwards the monopoles redshift as matter. This leads to a comoving number density
\be\label{eq:Monoscaling}
Y_M\approx \frac{\zeta}{4} \gamma_*^{-1/4} (r_M\MP)^{-3/2}\,,
\ee
see the green line labeled `global' in Fig.~\ref{fig:SOPT}.
In the intermediate case $\xi_{\rm KZ} <r_M<H(T_c)^{-1}$, the field configuration after the phase transition still consists of many overlapping monopoles which will efficiently annihilate. Assuming that this takes place on a short timescale, the effective correlation length 
is given by $\xi\approx r_M$, and the comoving number density becomes
\be\label{eq:Monoann}
Y_M\approx \dfrac 14 \gamma_*^{-1}  (r_M T_c)^{-3} \,,
\ee
which interpolates between the KZ and global limits, see Fig.~\ref{fig:SOPT}. In all cases the abundance is further reduced by monopole annihilations, see section \ref{sec:MonoAnn}.

\begin{figure}[tb!]
\begin{center}
    \includegraphics[width=.45\textwidth]{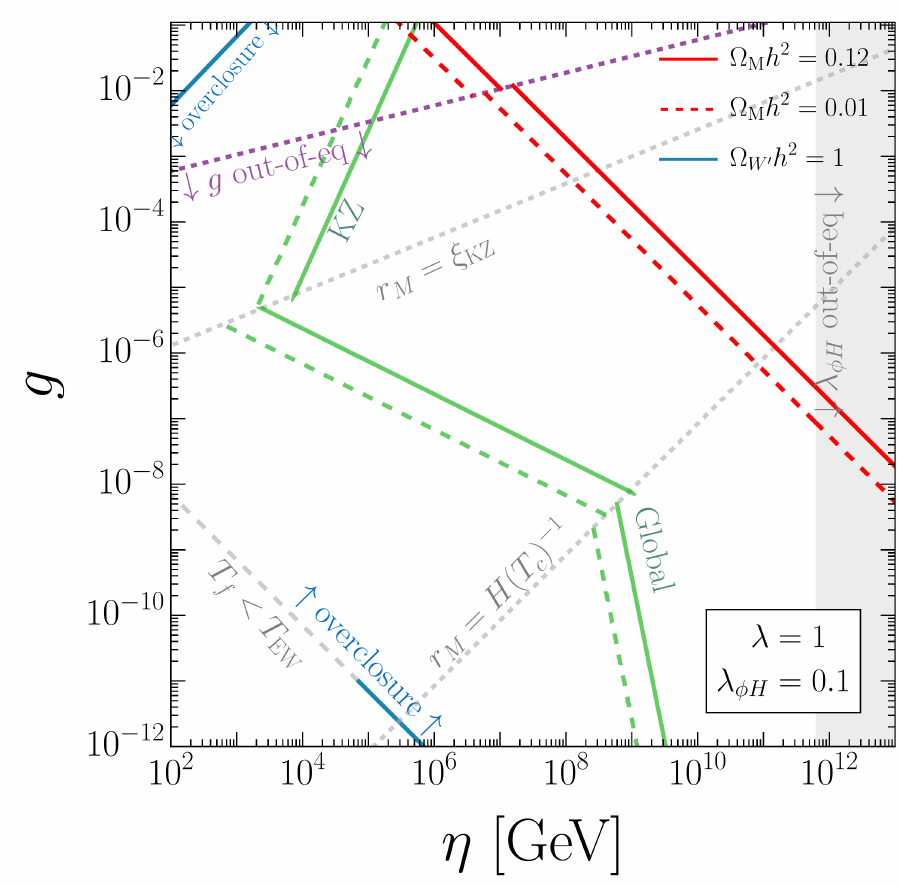}
    \caption{Relic density of the vectors $W'$ (blue) and of the monopoles $M$, just after a SOPT (green) and today (red).
    In the grey region the whole dark sector is not thermalised with the SM;
    below the purple line, the transverse dark gauge bosons are not thermalised with the rest of the dark sector. 
    The $W'$ overclosure abundance is determined by (non-)relativistic freeze-out in the (top-left) bottom-left corner; we neglected EW symmetry breaking, which would further enhance $\Omega_{W'}$ for $T_f < T_{\text{EW}}$. For $r_M< \xi_{\rm KZ}$, the monopoles are produced by the KZ mechanism, for $r_M>1/H(T_c)$ they are produced as global monopoles, while for the intermediate region we took  $\xi\approx r_M$; in all three cases, later monopole annihilations control the final abundance.
    }
\label{fig:SOPT}  
\end{center}
\end{figure}

\section{Monopoles from first-order phase transitions}\label{sec:MonoFOPT}

The hallmark of a FOPT is the presence of two degenerate minima of the thermal effective potential at the critical temperature, separated by a barrier. 
The scalar field remains trapped in the symmetry-preserving false vacuum as the universe cools down. At the nucleation temperature, $T_n$, quantum tunneling or thermal fluctuations are efficient enough for the phase transition to proceed, via nucleation of bubbles of true vacuum in a background of metastable phase. 

These bubbles expand and eventually percolate. The correlation length $\xi$ at the end of the phase transition is given by the average bubble radius at percolation $R_p$, such that the comoving monopole number density is
\be\label{eq:nMFOPT}
Y_M\approx \frac{1}{32} (\gamma_*^{\rm reh})^{-1}\left( R_p\, T_{\rm reh}\right)^{-3}\,.
\ee
Here $T_{\rm reh}$ is the temperature to which the universe is reheated by bubble collisions.
The value of $R_p$ strongly depends on the phase transition parameters, which control bubble nucleation and expansion dynamics. In the Appendix~\ref{Rp} we review these subjects in more detail and show how to compute $R_p$ in the various regimes of interest; below we summarise the main results.

\begin{figure}[tb!]
\begin{center}
\includegraphics[width=.45\textwidth]{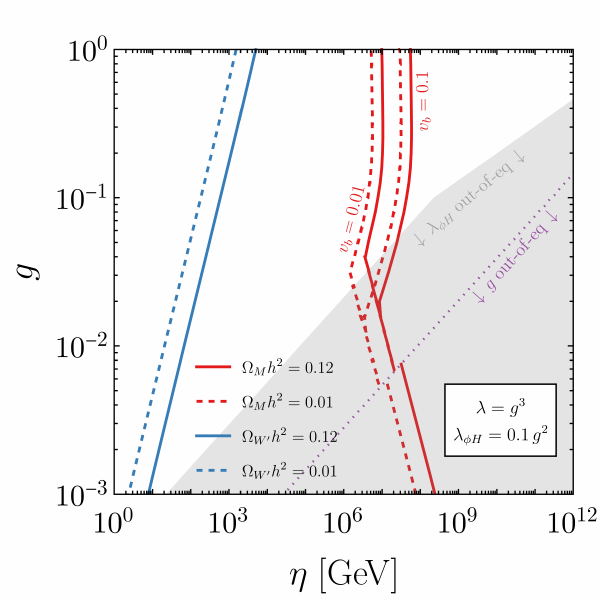}
    \caption{The monopole abundance (red) after a wFOPT, taking into account later annihilations which become relevant for $g\lesssim 2\times 10^{-2}$.
    The corresponding $W'$ abundance (blue) 
    is determined by standard freeze-out, driven by annihilations into dark photons. 
    Grey region and purple line are defined as in Fig.~\ref{fig:SOPT}.
   }
\label{fig:wFOPT}  
\end{center}
\end{figure}

A phase transition is weakly first-order (wFOPT) if it occurs when the universe is still radiation dominated, so that $T_{\rm reh}\approx T_p$. In this case the mean bubble radius at percolation is given by
\be
    R_p^{(w)} = \left(\frac{8\pi v_b^3}{0.29}\right)^{1/3}\beta^{-1}\,,
\label{rpw}\ee
where $v_b\ll 1$ is the terminal bubble wall velocity. The rate of completion of the phase transition $\beta$ is defined in terms of the bubble nucleation rate per unit volume,  $\Gamma(T) \approx T^4 e^{-S_3/T}$, where $S_3$ is the euclidean action of the bubble, according to
\be\label{eq:betaMAIN}
    \beta \equiv \frac{d\log \Gamma}{d t}\bigg\rvert_{t_p} 
    \simeq H(T_p) T_p \frac{d}{d T}\left(\dfrac{S_3}{T}\right)\bigg\rvert_{T_p}\,.
\ee
By numerically evaluating $S_3$ and $T_p$, one can use Eq.~\eqref{eq:nMFOPT} to obtain the
monopole abundance, illustrated by the red lines in Fig.~\ref{fig:wFOPT} for two different values of $v_b$.

\begin{figure}[tb!]
\begin{center}
\includegraphics[width=.45\textwidth]{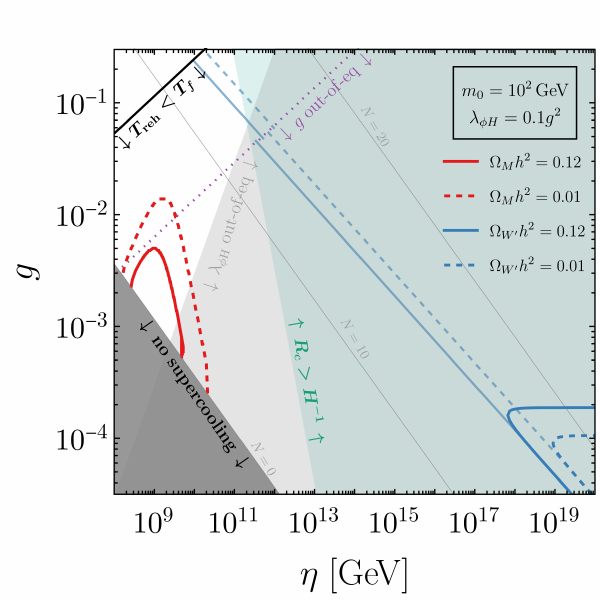}\quad
\includegraphics[width=.45\textwidth]{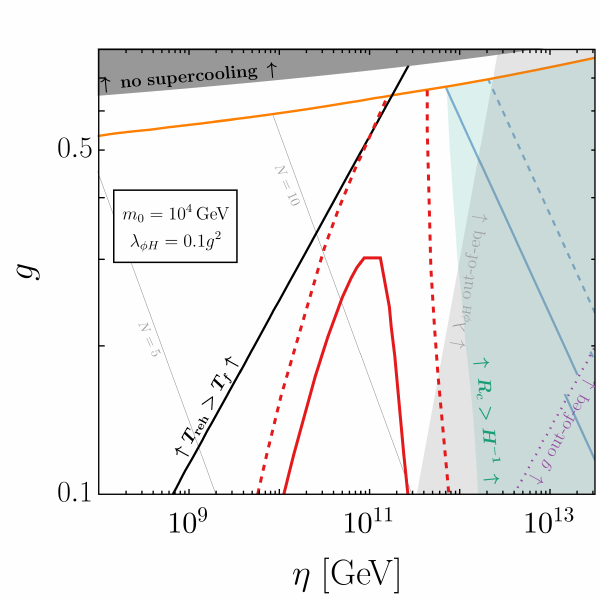}
    \caption{ Relic abundance of monopoles (red) and dark gauge bosons (thick blue), for a supercooled sFOPT, followed by instantaneous reheating. Thin blue lines show the $W'$ abundance at the end of supercooling, neglecting the sub-thermal population generated after reheating. 
    Grey lines show isocontours of the number $N$ of $e$-folds of thermal inflation, which dilute the $W'$s. Above the orange line, nucleation occurs independently of $m_0$ at much larger temperatures so that $N$ sharply decreases.
    Above the black line, the universe is reheated to a temperature large enough to restore $W'$ thermal equilibrium. 
    Inside the dark grey region the phase transition completes before the universe enters vacuum domination.
    The green shading indicates the region where the effect of gravity on the tunneling rate should be taken into account. }
\label{fig:MonoWAbunsFOPT}  
\end{center}
\end{figure}

A phase transition is strongly first-order (sFOPT) if the universe becomes vacuum-dominated before bubbles start nucleating, so that there is substantial supercooling.
If supercooling ends by the potential barrier disappearing at some temperature $T>0$, the phase transition can be a fast sFOPT. An example is given by a deformed Coleman-Weinberg scenario with a small negative $V_{\rm eff}''(0)\equiv -m_0^2$ at zero temperature. In this case, the critical bubble radius $R_c$ of bubbles at nucleation may be sizeable and should be added to the bubble radius gained from expansion, in order to estimate the final radius at percolation,
$R_p = R_c + R_{\rm exp}$. Here $R_{\rm exp}$ is still given by the expression in Eq.~\eqref{rpw}, now with $v_b\approx 1$.
In the thick-wall approximation, in terms of the effective potential parameters of Eq.~\eqref{eq:Veff}, one has 
\be
R_c^2=\frac{1}{m^2(T_n)}\frac{3\kappa(T_n)}{3\kappa(T_n)-1+\sqrt{1-4\kappa(T_n)}}\,,
\ee
with $\kappa(T_n) \equiv \lambda(T_n) m^2(T_n)/\delta^2(T_n)$.
The red lines in Fig.~\ref{fig:MonoWAbunsFOPT} show the resulting monopole abundance created by a sFOPT, for two different values of $m_0$. The radius at percolation is dominated by $R_{\rm exp}$ for small values of $\eta$ (rising red curves) and by $R_c$ for large values (falling red curves).

\section{Monopole annihilation}\label{sec:MonoAnn}

The monopole  number density produced by the phase transition may be reduced by monopole-antimonopole annihilation \cite{Zeldovich:1978wj,Preskill:1979zi} (see \cite{Vilenkin:2000jqa} for a review). Monopoles can dissipate their energy by moving through a plasma of relativistic charged particles, in our case the $W'$ gauge bosons.
This allows the formation of monopole-antimonopole bound states, which eventually annihilate. The diffusive capture process takes effect as long as the mean free path of the monopole in the plasma is smaller than the capture radius. It stops once the monopole abundance is sufficiently diluted, or when the $W'$ bosons become non-relativistic, whichever happens first. It may also happen that the monopole density at production was never large enough to allow for efficient annihilation in the first place. When annihilation is relevant, the final  yield is \cite{Preskill:1979zi}
\begin{equation}\label{eq:YAnn}
    Y_M = \frac{\mathcal{B} g^2}{16\pi}\gamma_*^{-1/2}\frac{T_a}{\MP},
\end{equation}
with $T_a$ the temperature where annihilation ceases to be effective, and $\mathcal{B}=g_{W'}\zeta(3)/\pi^2$ for a plasma of $W'$ bosons.

\section{Vector boson abundance}

The $W'$ vector bosons, as the lightest (and only) particles carrying dark electric charge, are also stable DM candidates. The abundance of transversely polarized $W'$ bosons and dark photons $\gamma'$ may be parametrically small in the limit of small $g$. The relevant process is $W'-\phi$ scattering before the dark-sector phase transition, with
\begin{equation}
    \langle \sigma v \rangle_{W'W'\rightarrow \phi\phi}=\frac{41 g^4}{4608\pi T^2} \,.
\end{equation}
The transverse polarisations of the three dark gauge bosons are not thermalised at the phase transition if $n_{W'}^{\rm eq}\langle \sigma v\rangle < H$, which requires  $T_c\gtrsim g^4 \left(1.2\times10^{15}\,{\rm GeV}\right)$. 
However, by assumption $\phi$ is thermalized with the SM via the Higgs portal, and therefore there is always a thermal population of dark Higgs bosons and of longitudinally polarized $W'$s, which are Nambu-Goldstone bosons in the global limit. From the thermally averaged cross-section for (massless) $\phi$ -- Higgs boson scattering,
\be
   \langle \sigma v\rangle_{\phi\phi \to HH} = \frac{\lambda_{\phi H}^2}{128\pi T^2}\,,
\ee
one deduces that the two sectors are thermalised at the time of the phase transition, $n^{\rm eq}_\phi \langle\sigma v\rangle>H$, provided that $T_c\lesssim \lambda_{\phi H}^2 \left(10^{14}\,{\rm GeV}\right)$.

In the case of a SOPT or a wFOPT, the present-day $W'$ abundance is determined by ordinary freeze-out, with different possible annihilation channels. Detailed expressions are provided in Appendix \ref{WFO}. The resulting $W'$ abundance is shown by the blue curves in Figs.~\ref{fig:SOPT} and \ref{fig:wFOPT}.

In contrast, in the case of a sFOPT with supercooling, the thermal population of massless $W'$s is diluted, before the phase transition, by a phase of thermal inflation. After the transition however, if the reheating temperature $T_{\rm reh}$ is much larger (not much smaller) than the $W'$ freeze-out temperature $T_f$, then the $W'$ will (partially) thermalise again, see Appendix~\ref{WFO} for explicit expressions. The resulting $W'$ abundance is shown in Fig.~\ref{fig:MonoWAbunsFOPT} by thick (thin) blue lines, assuming instantaneous reheating (suppressing the reheating contribution).

\section{Conclusions and perspectives}\label{perspectives}

We charted the magnetic monopole relic abundance across the parameter space of a minimal 't Hooft-Polyakov dark sector,
which is fully characterised by three parameters: the symmetry-breaking scale $\eta$, the (electric) gauge coupling $g$, and the scalar self-coupling $\lambda$ (or, equivalently, a scalar mass parameter $m_0$). 
Our analysis shows that $\Omega_M \ll \Omega_{W'}$ for all possible (perturbative) ranges for $g$, $\lambda$ and $\eta$. In particular, $\Omega_{W'}$ can be parametrically small in the limit $m_{W'}\to 0$, however in this limit $\Omega_M$ is even more suppressed, see Fig.~\ref{fig:SOPT}. In addition, while a period of thermal inflation during a supercooled FOPT may significantly reduce the $W'$ relic density, $W'$s may then be recreated after reheating. Even if their resulting abundance is subthermal, they still overclose the universe by far in the parameter space region where $\Omega_M\sim 0.1$, see Fig.~\ref{fig:MonoWAbunsFOPT}.
To conclude, the DM relic density is always dominated by the lightest electrically-charged particle, and never by the lightest magnetically-charged state.

Comparing with the previous literature, we agree with the prediction of $\Omega_M$ for a SOPT  found by \cite{Murayama:2009nj}. However, this work did not study monopoles embedded in a complete, realistic model, such as the 't Hooft-Polyakov model including $W'$ bosons. We agree with \cite{Baek:2013dwa} that, in the specific region with $\Omega_{W'}\sim 0.1$ and a SOPT,  $\Omega_M$ is negligible.  Finally, we disagree with  \cite{Khoze:2014woa} that monopoles can account for an ${\cal O}(1)$ fraction of the observed DM relic density.\footnote{Most significantly, we find that the monopole abundance after annihilation resulting from Eq.~\eqref{eq:MonoSOPT} scales differently with $g$ than Eq.~(3.39) of the arXiv version 3 of \cite{Khoze:2014woa}.}

The negative conclusion for monopole DM may change if one relaxes some of our assumptions:
\begin{itemize}
\item There are potentially large corrections when the dark-sector couplings become non-perturbative, or when gravity becomes significant for the phase-transition tunnelling process.
\item If the Higgs portal $\lambda_{\phi H}$ (or any alternative interaction between the dark sector and the SM) is too small to maintain thermalisation, the dark-sector temperature $T'$ can be smaller than $T$ and the ratio $\Omega_M/\Omega_{W'}$ may vary with $T'/T$. On the other hand, if $\lambda_{\phi H}$ is comparable to or larger than both $\lambda$ and $g^2$, then the dark phase transition can be significantly deformed, and interfere with the electroweak phase transition. In particular, the role of $m_0^2$ in triggering a sFOPT could be taken by $\lambda_{\phi H}v^2$ with $v$ the Higgs vacuum expectation value.
\item The dark sector could be extended: (i) light charged particles could allow $W'$ to decay; (ii) charged scalars other than $\phi$ could determine different spontaneous symmetry-breaking patterns, with other kinds of monopoles and additional topological defects; (iii) $\SO{3}$ can be replaced by a larger gauge group.
\end{itemize}
In a separate paper \cite{Brummer:2026wma}, we show that some of these extensions allow for monopoles to naturally dominate the DM abundance, with a specific associated phenomenology.

\subsection*{Acknowledgments}

We thank Marco Cirelli, Yann Gouttenoire, Yu Hamada, Valya Khoze, Jean-Loïc Kneur, and Rudnei Ramos for useful discussions and correspondence. This project has received funding from the European Union’s Horizon Europe research and innovation programme under the Marie Sklodowska-Curie Staff Exchange grant agreement No 101086085 – ASYMMETRY.\\

\subsection*{Data availability}

The plots appearing in this article can be found at \cite{FigShare}.
\newpage

\onecolumngrid
\makeatletter
\long\def\@makefntext#1{%
  \parindent 1em%
  \noindent
  \hb@xt@1.8em{\hss\@makefnmark}\parbox[t]{\textwidth}{#1}}
\makeatother  
\appendix

\section{Derivation of the bubble radius at percolation}\label{Rp}

We have shown, in the main text, that the comoving number density of monopoles produced during a FOPT can be estimated as
\be\label{eq:nMFOPT}
Y_M\approx \frac{1}{32} (\gamma_*^{\rm reh})^{-1}\left( R_p\, T_{\rm reh}\right)^{-3}\,.
\ee
Determining  precisely  the bubble radius at percolation, $R_p$, would require numerical simulations of the bubble evolution. However, here we will use semi-analytical approximations which can be obtained in the limiting cases of interest. To this end, we now collect some results from the literature regarding bubble nucleation and expansion in cosmology; for details see the recent review \cite{Athron:2023xlk}.

Neglecting quantum tunneling, the thermal bubble nucleation rate per unit volume is given by \cite{Linde:1981zj}
\be\label{eq:Gammav}
    \Gamma(T) \approx T^4 e^{-S_3/T},
\ee
where $S_3$ is the euclidean action of a bubble, evaluated along the O$(3)$-symmetric bounce solution describing finite-temperature transitions. 
The \textit{nucleation temperature} $T_n$ is defined as the temperature at which there is, on average, one bubble per Hubble volume\footnote{ In principle, one should integrate the probability of nucleation per Hubble volume over time. However, the integral is dominated by the nucleation time $t_n$ and Eq.~\eqref{eq:Tn} provides a good estimate for $T_n$.}
\be\label{eq:Tn}
 \Gamma(T_n) = H(T_n)^4\,,
\ee
signalling the onset of the phase transition.

The latent heat parameter $\alpha$ is defined as the ratio between vacuum and radiation energy densities at the moment of nucleation:
\be\label{eq:alpha}
    \alpha \equiv \frac{\Delta V(T_n)}{\rho_r(T_n)} = \frac{\Delta V(T_n)}{3\gamma_*\,T_n^4}\,.
\ee
 If $T_n \ll T_c$, the vacuum energy density eventually comes to dominate before the first bubbles start nucleating. This happens at $T=T_{\rm eq}$, which is implicitly defined by
\be
    T_{\rm eq} \equiv \left(\frac{\Delta V (T_{\rm eq})}{3\gamma_*}\right)^{1/4}\,.
\ee
Approximating $\Delta V(T_{\rm eq})\approx\Delta V(T_n)\approx \Delta V$, with $\Delta V$ the vacuum energy difference at zero temperature, gives $\alpha \approx (T_{\rm eq}/T_n)^4$. If $\alpha >1$, bubbles nucleate in a vacuum-dominated universe, and a \textit{strongly first order phase transition} (sFOPT) is realised. Otherwise, the phase transition takes place during radiation domination. This is the case for a \textit{weakly first order phase transition} (wFOPT).

The phase transition completes at the percolation time $t_p$ at temperature $T_p$.
We define the completion rate $\beta$ by
\be\label{eq:beta}
    \beta \equiv \frac{d\log \Gamma}{d t}\bigg\rvert_{t_p} = -H(T_p) T_p \frac{d \log \Gamma}{d T}\bigg\rvert_{T_p} \simeq H(T_p)T_p\frac{d}{dT}\left(\frac{S_3}{T}\right)\bigg\rvert_{T_p}\,.
    \ee
In the following, we will dub as \textit{fast} those phase transitions for which $\beta/H(T_p) \gg 1$. Generally wFOPTs are fast, while sFOPTs may or may not be fast.

Finally, we define $\mathcal{P}_{f}(T)$ to be the probability of finding one point in space in the false vacuum, at any given time or temperature; and $I(T)\equiv -\log \mathcal{P}_{f}(T)$. The continuum percolation threshold for spherical objects in three dimensions is $29\%$, so the phase transition ends when $\mathcal{P}_{f}(T_p)=0.71$ and $I(T_p)=0.34$. If the initial bubble size at nucleation is negligible with respect to the size gained during the expansion, and if the Hubble parameter can be regarded as constant during the expansion process (because the universe is vacuum dominated or the phase transition is fast), then $I(T)$ may be approximated as
\be\label{eq:I(t)}
     I(T) \approx 8\pi v_b^3\frac{\Gamma(T)}{\beta^4}\,.
\ee
Here $v_b$ is the terminal bubble wall velocity.

\begin{figure}[tb!]
\begin{center}
\includegraphics[width=.4\textwidth]{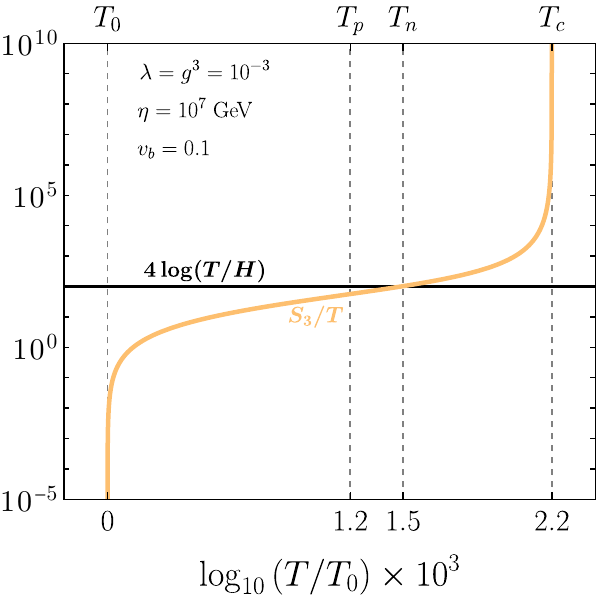}\qquad\includegraphics[width=.4 \textwidth]{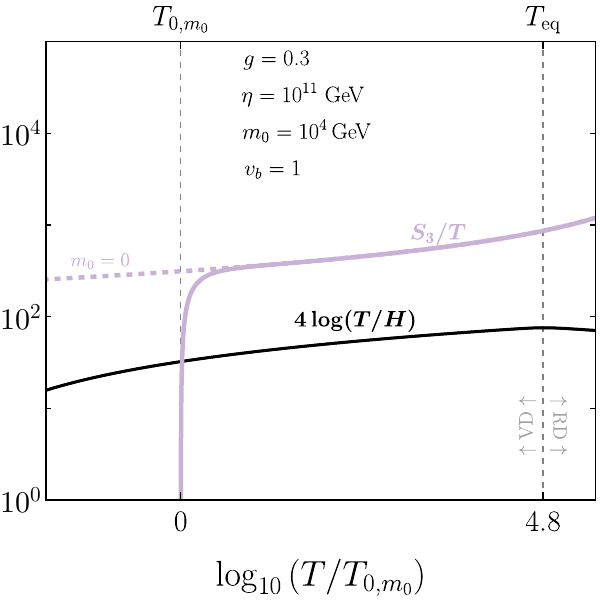}
    \caption{\textit{Left panel:} Evolution of the bounce action $S_3$ (orange) as a function of the temperature, in the case of a wFOPT. The nucleation temperature $T_n$ corresponds to the crossing of the orange and black lines. The phase transition ends at $T=T_p$, where bubbles of the true vacuum percolate. $T_0$ is the temperature at which the symmetry-preserving point ceases to be a local minimum, and the thermal barrier vanishes. \textit{Right panel:} Bounce action $S_3$ (purple) for a supercooled sFOPT, in a mass-deformed Coleman-Weinberg model. Driven by the presence of a small mass parameter $m_0$, the phase transition takes place at a temperature $T_{0,m_0}\ll T_c$ where the barrier disappears.  Bubbles start nucleating and percolate very close to this temperature: $T_n\approx T_p \approx T_{0,m_0}$. In the Coleman-Weinberg case ($m_0 = 0$), the Universe would remain stuck in the false vacuum (dashed purple). Vacuum (radiation) dominates the energy budget to the left (right) of the $T_{\rm eq}$ line.}
\label{fig:FOPT}  
\end{center} 
\end{figure}

The bounce action can be obtained \cite{Levi:2022bzt} from the thermal effective potential, defined in the main text, with its coefficients computed in the high-temperature expansion. This allows to numerically evaluate quantities such as $I(T)$ and $\beta$ which will enter into the monopole abundance. Fig.~\ref{fig:FOPT} illustrates the behavior of $\log\Gamma/T^4\approx -S_3/T$ as a function of the temperature in a fast wFOPT and in a fast, supercooled sFOPT. Note the different scales on the horizontal axes: indeed $\beta$, which roughly corresponds to the slope of the $S_3/T$ curve as it crosses the black line, is large in both cases, hence both transitions are fast.

To proceed further, we distinguish several cases:
\begin{itemize}
 \item
In the case of a (fast) wFOPT, neglecting the expansion of the universe between $T_n$ and $T_p$ gives the bubble number density \be\label{eq:nbwFOPT}
     n_b^{(w)}(T) = \frac{\Gamma(T)}{\beta I(T)}\left[1-\mathcal{P}_f(T)\right]
\ee
and thus, with Eq.~\eqref{eq:I(t)} and $R_p=\left[n_b(T_p)\right]^{-1/3}$, the mean bubble radius at percolation
\be\label{eq:Rpw}
    R_p^{(w)} = \left[\frac{8\pi v_b^3}{1-\mathcal{P}_f(T_p)}\right]^{1/3}\beta^{-1}\,.
\ee
The final result for the comoving monopole number density is then, from Eq.~\eqref{eq:nMFOPT},
\be\label{eq:nMw}
   Y_M^{(w)} = \frac{1-\mathcal{P}_f(T_p)}{256\pi \,\gamma_*\,v_b^3}\frac{\beta^3}{T_p^3}\,.
\ee
Here we have taken $T_{\rm reh} = T_p$,
since for a wFOPT the universe remains radiation dominated and its evolution is approximately adiabatic. The bubble velocity is treated as a free parameter. On general grounds, we expect the radiation bath to exert a non-negligible friction on bubble walls, so that $v_b\ll 1$.  Eq.~\eqref{eq:nMw} can now be evaluated numerically.
The parameter space for monopole DM produced by a wFOPT is illustrated by the red lines in Fig.~3 in the main text, for two different values of $v_b$.

\item
If the phase transition is a slow sFOPT, the expansion of the universe during bubble evolution is no longer negligible. However, an analytical approximate result for the bubble number density can still be obtained in a de Sitter background \cite{Levi:2022bzt}:
\be
    n_b^{(s)}(T) = \frac{\Gamma(T)}{\beta}I(T)^{-1-3\frac{H}{\beta}}\left[\tilde{\Gamma}\left(1+3\frac{H}{\beta}, 0\right)-\tilde{\Gamma}\left(1+3\frac{H}{\beta},I(T)\right)\right]\,,
\ee
where $\tilde{\Gamma}$ is the incomplete gamma function. We may set $v_b=1$ since the bubble walls are no longer subject to friction from the radiation bath, which has been diluted by the exponential expansion.  Following the same steps as above, one obtains the comoving monopole number density. In a sFOPT, the high-temperature expansion of the thermal effective potential is generally unreliable for field excursions all the way to the true vacuum. However, if there is sufficient supercooling, $T_n\ll\eta$, it can still be used for modelling the tunneling process \cite{Salvio:2023qgb,Salvio:2023ynn}. Note that, for a sFOPT, we set $T_{\rm reh}=T_{\rm eq}$ as the universe is dominated by the false-vacuum energy before reheating. In this regime, the monopole abundance turns out to be much smaller than the $W'$ abundance. The reason is twofold: slow sFOPTs are characterised by small values of $\beta$, compared to fast ones, and consequently by a smaller production of monopoles, cfr.~Eq.~\eqref{eq:nMw}; moreover, the stage of supercooling undergone by the universe is shorter with respect to fast sFOPTs, resulting into a milder dilution of the $W'$ abundance, see discussion in sec.~\ref{WFO}. Therefore, we do not illustrate further this intermediate region of parameter space in the main text.

\item
In models of radiative symmetry breaking with $V_{\rm eff}''(0)\geq 0$, where $V_{\rm eff}$ is the zero-temperature effective potential, a barrier persists down to arbitrarily low temperatures. This is the case e.g.~in the pure Coleman-Weinberg scenario. The universe may never be able to thermally tunnel to the true vacuum (but quantum fluctuations, which we have ignored, may still trigger the phase transition  \cite{DelleRose:2019pgi,Lewicki:2021xku}).

If, instead, supercooling ends by the barrier disappearing at some temperature $T>0$, the phase transition can be a fast sFOPT, provided that it still proceeds via tunnelling rather than classical rolling. An example is given by a deformed Coleman-Weinberg scenario with a small negative $V_{\rm eff}''(0)\equiv -m_0^2$. In this case, however, it turns out that the critical bubble radius $R_c$ of bubbles at nucleation may no longer be negligible; in other words, the final bubble size can be dominated by the initial bubble size and not by bubble expansion. Therefore, Eq.~\eqref{eq:I(t)} no longer holds. Instead we obtain a lower bound on the bubble radius at percolation from $R_c$. In the thick-wall approximation, in terms of the effective potential parameters defined in the main text, one has
\be
R_c^2=\frac{1}{m^2(T_n)}\frac{3\kappa(T_n)}{3\kappa(T_n)-1+\sqrt{1-4\kappa(T_n)}}\,,
\ee
with $\kappa(T_n) \equiv \lambda(T_n) m^2(T_n)/\delta^2(T_n)$.
A bound on the correlation length is then obtained as
\be
\xi\lesssim R_{\rm exp}+R_c
\ee
where $R_{\rm exp}$ is the radius to which a bubble of negligible initial size would have expanded until percolation occurs. For a fast sFOPT, this is given by Eq.~\eqref{eq:Rpw}, with $v_b = 1$.
The red lines in Fig.~4 in the main text show the resulting monopole abundance created by a sFOPT with supercooling, for two different values of $m_0$. 
\end{itemize}

\section{Vector boson abundance from freeze-out or from supercooling}\label{WFO}

Assuming a thermal population of $W'$ vector bosons after the dark-sector phase transition, the present-day $W'$ abundance is determined by 
freeze-out.
If they freeze out when they are non-relativistic ($x_f\equiv m_{W'}/T_f \gg 1$, where $T_f$ is the freeze-out temperature), their abundance is given by
\be\label{eq:VectDM}
    \Omega_{W'}h^2 = 1.1\times 10^{-9}\left(\gamma_*^f\right)^{-1/2}\left(\frac{x_f}{20}\right)\left[\langle\sigma v\rangle\;{\rm GeV}^2\right]^{-1}\,.
\ee
where the value of $x_f$ depends logarithmically on the cross-section \cite{Kolb:1990vq}.
If they are relativistic at freeze-out ($x_f\ll 1$), their abundance is independent of the annihilation cross-section and given by 
\begin{equation}
    \Omega_{W'} h^2 = 0.12
    \dfrac{g_{W'}}{6}\left(\gamma_*^f\right)^{-1}\dfrac{m_{W'}}{2.39\ {\rm eV}}\,.\label{hotDM}
\end{equation}

Note $g_{W'}=6$ if all three polarisations are thermalised, but $g_{W'}=2$
if only longitudinal $W'$s are in the bath. 
These dark gauge bosons would overclose the Universe for masses above $\sim 100$ eV, therefore they cannot constitute cold DM. Sub-keV $W'$s are subject to strong constraints on hot DM, as well as to bounds on dark radiation.

The massive $W'$s annihilate predominantly into either dark photons, dark Higgs bosons or SM particles, depending on the couplings. For $\lambda_{\phi H}$ sufficiently small and $\lambda> g^2$, the dominant process is $W'^+W'^-\to \gamma'\gamma'$, since the annihilation into $\rho\rho$ is kinematically forbidden. The thermally averaged cross-section is well approximated by its s-wave component,
\begin{equation}
    \langle \sigma v \rangle_{W'^+W'^-\to \gamma'\gamma'} \approx \frac{19 g^4}{72\pi m_{W'}^2}\,.
\label{photonFO}\end{equation} 
This process controls the $W'$ non-relativistic freeze-out in the region $\lambda\gg g^2$, corresponding to a SOPT, see Fig.~\ref{fig:SOPT}.
\begin{figure}[tb!]
\begin{center}
\includegraphics[width=.45\textwidth]{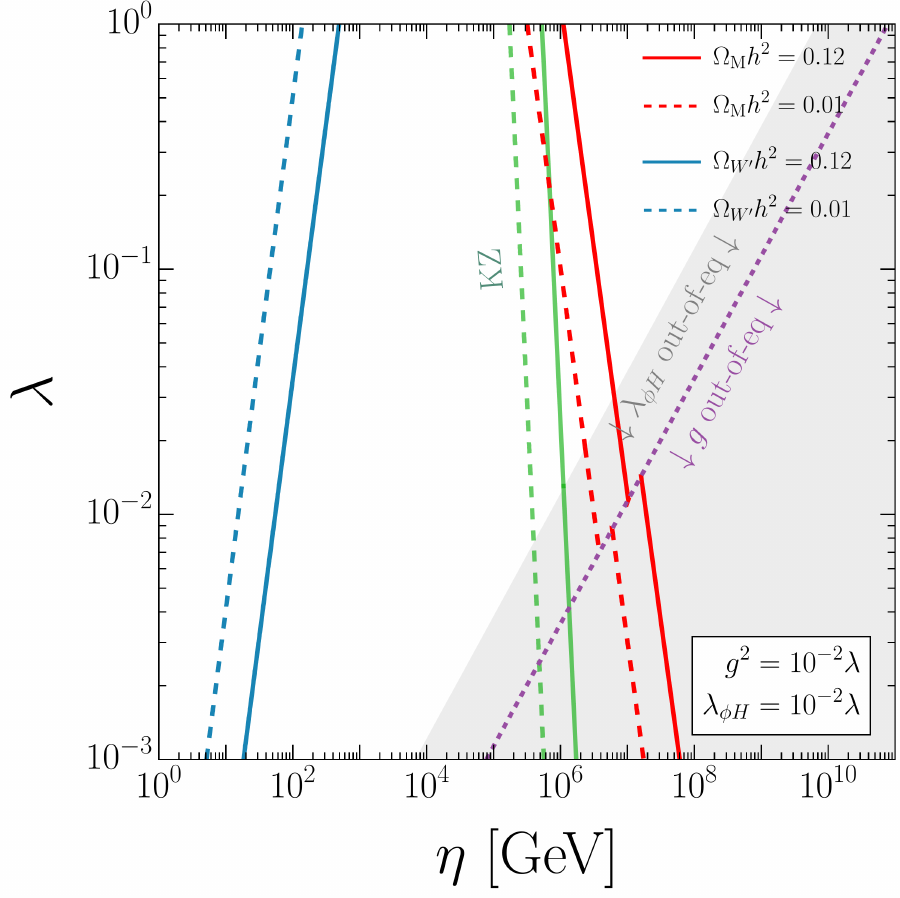}
    \caption{Relic density of the vectors $W'$ (blue) and of the monopoles $M$ (red) after a SOPT, as a function of the scalar self-interaction $\lambda$ and the dark spontaneous-symmetry-breaking scale $\eta$, with the other couplings fixed as indicated in the legend.
    The relevant process setting the $W'$ abundance is non-relativistic freeze-out driven by annihilations into dark photons, cfr.~Eq.~\eqref{photonFO}. The monopoles are produced by the Kibble-Zurek mechanism (green) and later undergo annihilations (red).  In the grey region the whole dark sector is not thermalised with the SM;
    below the purple line, the transverse dark gauge bosons are not thermalised with the rest of the dark sector.}
\label{fig:SOPT}  
\end{center}
\end{figure}
For $\lambda/g^2\lesssim 1$, also the annihilation into dark scalars becomes relevant, as $m_\rho\lesssim m_{W'}$. 
Taking again the s-wave approximation, we find that the first terms in the $\lambda/g^2$ expansion are
\begin{equation}
\langle \sigma v \rangle_{W'^+W'^-\to \rho\rho} \approx \frac{11g^4}{144\pi m_{W'}^2}-\frac{14g^2\lambda}{192\pi m_{W'}^2}+\frac{31\lambda^2}{2304\pi m_{W'}^2} \,.
\label{rhoFO}\end{equation} 
This result agrees with \cite{2925534} up to a misprint in one of the vertices.
In the region $g^2\gtrsim \lambda$, corresponding to a FOPT, the non-relativistic freeze-out is controlled by the sum of cross-sections into dark photons and scalars (our result for such sum differs from the one reported in \cite{Khoze:2014woa}). Fig.~3 in the main text shows the resulting $W'$ abundance for the wFOPT regime. 

Coming to annihilation into SM particles through the Higgs portal, the relevant process is $W'W'\rightarrow HH$, as long as the freeze-out occurs before the EW symmetry-breaking scale, $T_f>T_{\rm EW}$. Since this annihilation is mediated by a $\rho$, in the limit $m_\rho\gg m_{W'}$ it turns out that the annihilation rate is so small that the $W'$s freeze-out when they are still relativistic. This is relevant in the region $\lambda \sim \lambda_{\phi H}\gg g^2$, with a $W'$ abundance given by Eq.~\eqref{hotDM}, see  Fig.~2 in the main text.
In the opposite limit, $m_\rho\ll m_{W'}$, the thermally averaged cross-section is not suppressed,
\begin{equation}
    \langle \sigma v \rangle_{W'W'\to HH}\approx \frac{\lambda_{\phi H}^2}{384\pi m_{W'}^2}\,, 
\label{WWHH}\end{equation}
and the freeze-out happens at $x_f \gg 1$.
In the region $g^2\sim \lambda_{\phi H} \gg \lambda$, this annihilation channel should be added to the one in Eqs.~\eqref{photonFO} and \eqref{rhoFO}, to determine the $W'$ freeze-out. 
However, to compute the initial monopole abundance, we 
neglected the Higgs portal corrections to the 
the $\phi$ thermal potential, therefore for consistency we assume that $\lambda$ and/or $g^2$ are significantly larger than $\lambda_{\phi H}$.

Let us notice that freeze-out may not be the relevant mechanism to determine the $W'$ abundance, when the $W'$ population is not thermal after the phase transition.
This happens in the case of a sFOPT with significant supercooling. 
Before the phase transition, the thermal population of massless $W'$s is diluted by a phase of thermal inflation, such that the abundance becomes \cite{Hambye:2018qjv}
\be\label{eq:WabunSupercool}
    \Omega_{W'}h^2 = 4.2\times 10^7 \left(\gamma_*^{\rm reh}\right)^{-1}\left(\frac{g_{W'}}{6}\right)\left(\frac{m_{W'}}{\rm GeV}\right) e^{-3N}\,.
\ee 
Here $N = \log \left(T_{\rm eq}/T_p\right)$ is the number of $e$-folds of supercooling. In addition, a substantial number of $W'$s can be created after reheating. If $T_{\rm reh}\gg T_f$, $W'$s will thermalize again and the dilution from supercooling will be obliterated. But even for $T_{\rm reh}\lesssim T_f$, a significant subthermal $W'$ population can be generated. Assuming instantaneous reheating, this latter contribution is given by \cite{Hambye:2018qjv}
\be
    \Omega_{W'}h^2 = 3.3\times 10^{23}\left(\gamma_*^{\rm reh}\right)^{-3/2} \left(\frac{g_{W'}}{2}\right)^2 m_{W'}^2 \langle \sigma v\rangle \left(1+2 \frac{m_{W'}}{T_{\rm reh}}\right)e^{-2 m_{W'}/T_{\rm reh}}.
\ee
This effect is relevant in our minimal model, where the ratio $m_{W'}/T_{\rm reh}$ is fixed and so the exponential suppression never leads to a parametrically small $\Omega_{W'}h^2$.

\end{document}